\documentclass
[superscriptaddress,secnumarabic,amssymb,amsmath,nobibnotes,aps,prd,showkeys,showpacs,nofootinbib,onecolumn]{revtex4}%
\usepackage{graphicx}
\usepackage{epstopdf}
\usepackage{epsf}
\usepackage{bm}
\usepackage{amsmath}
\usepackage{amsfonts}
\usepackage{amssymb}%
\providecommand{\U}[1]{\protect\rule{.1in}{.1in}}

\newcommand{\be}{\begin{equation}}
\newcommand{\ee}{\end{equation}}

\newcommand{\mincir}{\raise
-3.truept\hbox{\rlap{\hbox{$\sim$}}\raise4.truept\hbox{$<$}\ }}
\newcommand{\magcir}{\raise
-3.truept\hbox{\rlap{\hbox{$\sim$}}\raise4.truept\hbox{$>$}\ }}

\begin{document}
\title{Dynamics of a two scalar field cosmological model with phantom terms}
\author{Andronikos Paliathanasis}
\email{anpaliat@phys.uoa.gr}
\affiliation{Institute of Systems Science, Durban University of Technology, Durban 4000,
South Africa}
\affiliation{Instituto de Ciencias F\'{\i}sicas y Matem\'{a}ticas, Universidad Austral de
Chile, Valdivia 5090000, Chile}
\author{Genly Leon}
\email{genly.leon@ucn.cl}
\affiliation{Departamento de Matem\'{a}ticas, Universidad Cat\'{o}lica del Norte, Avda.
Angamos 0610, Casilla 1280 Antofagasta, Chile.}

\begin{abstract}
We perform a detailed analysis on the dynamics of a Chiral-like cosmological
model where the scalar fields can have negative kinetic terms. In particular,
we study the asymptotic dynamics for the gravitational field equations for
four different models in a spatially flat
Friedmann--Lema\^{\i}tre--Robertson--Walker background space. When one of the
scalar fields is phantom, we calculated that the cosmological fluid can
evolves such that the parameter for the equation of state crosses twice the
phantom divide line without the appearance of ghosts. Moreover, the
cosmological viability of these four models is discussed.

\end{abstract}
\date{\today}
\maketitle

\section{Introduction}

The detailed analysis of recent observations
\cite{Teg1,Teg2,Teg3,Teg4,Teg5,Teg6,Teg7,Teg8,Teg9,Teg10,Teg11,Teg12} supports
the idea that the universe has gone under two acceleration phases during its
evolution. An accelerated phase of expansion during its very early phase of
evolution, known as inflation \cite{guth,linde}, which occurred prior to the
radiation-dominated era; and a recently initiated accelerated phase of
expansion, known as late time cosmic acceleration, driven by the so-called
dark energy \cite{copeland,Clifton1,Nojiri:2017ncd,Ame10,Ame11}.

Within the framework of General Relativity, scalar fields play a significant
role in gravitational physics because they provide theoretical mechanisms for
the theoretical explanation of the observations. As inflaton is characterized
by the scalar field that is responsible for the early time acceleration phase
of the universe \cite{guth,linde}, the same model can provide dynamical terms
in the gravitational field equations with antigravitating behaviour and can be
used also as a model for the description of the acceleration of the late
universe. The most well-known scalar field model which has been studied in the
literature is the quintessence model \cite{Ratra,qq1,qq2,qq3,qq4,qq5,qq6}.
Alternative scalar-field models which have been proposed in the literature
are: the phantom fields \cite{ph1,ph2,ph3}, the quintom model
\cite{quin00,quin2,quin3,quin4,quin5,quin6}, the Chiral model
\cite{ch1,ch2,ch3,ch4}, the k-essence scalar field model \cite{kes1,kes2,kes3}%
, Galileon \cite{giampero} and many others \cite{hor1,hor2,hor3}.

An important common feature of the scalar-field models is that the
gravitational field equations are of second-order, as in the case of General
Relativity. Indeed, scalar fields introduce new degrees of freedom which are
necessary in order to provide the necessary behaviour of the dynamics.
Moreover these new degrees of freedom can attribute higher-order derivatives
which are introduced in gravitational physics by theories of gravity
alternatives of General Relativity \cite{so1}.

In this study we are interested in the study of the dynamics of a
generalization of the Chiral cosmological model. The Chiral model belongs to
the family of the multifield scalar-field theories, comprising two canonical
scalar fields which interact in the kinetic and in the potential parts. In
particular, the two scalar fields are involved in a two-dimensional space of
negative constant curvature and Euclidean signature. There are various studies
in the literature on this cosmological model \cite{ch1,ch2,ch3,ch4} with some
interesting results. It has been shown that it can describe the whole dark
sector of the universe, that is, it can been seen as a unified dark model for
the description of the dark energy and of the dark matter \cite{an1,an2}.
Moreover, it can provide two accelerated eras \cite{an3}, while, due to
quantum transitions in the early universe, the physical fluid can cross the
phantom divide line \cite{an4}. The latter is possible when the kinetic part
of one of the two fields changes sign and has a phantom behaviour where the
model can be seen as extension of the quintom theory where the one field is
quintessence and the second field is phantom.

The global dynamics and the cosmological eras provided by the Chiral cosmology
are studied in detail in \cite{an2}. \ In the following Sections we perform a
complete study on the asymptotic dynamics and behaviour for the cosmological
field equations of Chiral-like cosmological models where now at least one or
both of the scalar fields can be phantom and can have negative kinetic energy.
The mathematical tools that we apply for this work are based on the
$H-$normalization approach \cite{cop1}, where the field equations are written
in terms of dimensionless variables and are expressed in terms of a
algebraic-differential system of first-order \cite{gn1,gn2,gn3}. For the
latter dynamical system the stationary points are determined, where any
stationary point describe a specific exact solutions that is a specific era in
the cosmological history. The stability of the stationary points is also
investigated. Such an analysis provides important results of the viability of
the background equations of a proposed cosmological model
\cite{vb1,vb2,vb3,vb4}. The plan of the paper is as follows.

In Section \ref{sec2} we present the cosmological model of our analysis. It is
that of two-scalar fields which interact in the kinetic term, in particular
the kinetic part defines a space of constant curvature but not necessarily of
Euclidean signature. In Section \ref{sec3} we rewrite the gravitational field
equations by using dimensionless variables in the $H$-normalization approach.
The main analysis of this work is presented in Section \ref{sec4} in which we
investigate the existence of stationary points and the physical properties of
the exact solutions at the stationary points. Finally, in Section \ref{sec5}
we summarize and we draw our conclusions.

\section{Field equations}

\label{sec2}

For the Gravitational Theory of our consideration we consider a two-scalar
field model of Chiral-like with Action Integral%
\begin{equation}
S=\int\sqrt{-g}dx^{4}\left(  R-\frac{\varepsilon_{\phi}}{2}g^{\mu\nu}%
\nabla_{\mu}\phi\nabla_{\nu}\phi-\frac{\varepsilon_{\psi}}{2}g^{\mu\nu
}e^{\kappa\phi}\nabla_{\mu}\psi\nabla_{\nu}\psi-V\left(  \phi\right)  \right)
, \label{sp.01}%
\end{equation}
where for the background space we consider that of a spatially flat
Friedmann--Lema\^{\i}tre--Robertson--Walker (FLRW) space with line element%
\begin{equation}
ds^{2}=-dt^{2}+a^{2}\left(  t\right)  \left(  dx^{2}+dy^{2}+dz^{2}\right)  .
\label{sp.02}%
\end{equation}
Spacetime (\ref{sp.02}) admits a six-dimensional Killing algebra, which we
assume is inherited by the scalar fields $\phi\left(  x^{\nu}\right)
,~\psi\left(  x^{\nu}\right)  $. Thus $\phi=\phi\left(  t\right)  $ and
$\psi=\psi\left(  t\right)  $. Parameters $\left(  \varepsilon_{\phi
},\varepsilon_{\psi}\right)  $ are constants and take values $\left(
\varepsilon_{\phi},\varepsilon_{\psi}\right)  =\left(  \pm1,\pm1\right)  $, in
the case for which $\left(  \varepsilon_{\phi},\varepsilon_{\psi}\right)
=\left(  +1,+1\right)  $ the Chiral model is recovered, while in the
case for which \ $\left(  \varepsilon_{\phi},\varepsilon_{\psi}\right)
=\left(  +1,-1\right)  $ with $\kappa=0$ the Action Integral (\ref{sp.01})
becomes that of the quintom theory. The parameter $\kappa$ is related with the
curvature of the two-dimensional manifold defined by the kinetic parts of the
scalar fields, which has the line element%
\begin{equation}
ds^{2}\left(  \phi,\psi\right)  =\varepsilon_{\phi}d\phi^{2}+\varepsilon
_{\psi}e^{\kappa\phi}d\psi^{2} \label{sp.03}%
\end{equation}
and curvature $R_{\left(  \phi,\psi\right)  }^{\left(  2\right)  }=-\frac
{1}{2\varepsilon_{\phi}}\kappa^{2}$. The line element (\ref{sp.03}) describes
an Einstein space of constant curvature, where $R_{\left(  \phi,\psi\right)
}^{\left(  2\right)  }<0$ when $\varepsilon_{\phi}>0$ and $R_{\left(
\phi,\psi\right)  }^{\left(  2\right)  }>0$ for $\varepsilon_{\phi}<0$.

For the line element (\ref{sp.02}) and from (\ref{sp.01}) it follows that the
gravitational field equations are
\begin{equation}
-3H^{2}+\frac{\varepsilon_{\phi}}{2}\dot{\phi}^{2}+\frac{\varepsilon_{\psi}%
}{2}e^{\kappa\phi}\dot{\psi}^{2}+V\left(  \phi\right)  =0~, \label{sp.04}%
\end{equation}%
\begin{equation}
2\dot{H}+3H^{2}+\frac{\varepsilon_{\phi}}{2}\dot{\phi}^{2}+\frac
{\varepsilon_{\psi}}{2}e^{\kappa\phi}\dot{\psi}^{2}-V\left(  \phi\right)  =0~,
\label{sp.05}%
\end{equation}%
\begin{equation}
\varepsilon_{\phi}\left(  \ddot{\phi}+3H\dot{\phi}\right)  -\kappa
\frac{\varepsilon_{\psi}}{2}e^{\kappa\phi}\dot{\psi}^{2}+V_{,\phi}\left(
\phi\right)  =0~ \label{sp.06}%
\end{equation}
and%
\begin{equation}
\ddot{\psi}+3H\dot{\psi}+\kappa\dot{\phi}\dot{\psi}=0~, \label{sp.07}%
\end{equation}
where $H=\frac{\dot{a}}{a}$ is the Hubble function. The energy density
$\rho_{eff}$ ~and the pressure $p_{eff}~$of the effective fluid source are
defined as follows
\begin{equation}
\rho_{eff}=\frac{\varepsilon_{\phi}}{2}\dot{\phi}^{2}+\frac{\varepsilon_{\psi
}}{2}e^{\kappa\phi}\dot{\psi}^{2}+V\left(  \phi\right)  ~,~ \label{sp.08}%
\end{equation}%
\begin{equation}
p_{eff}=\frac{\varepsilon_{\phi}}{2}\dot{\phi}^{2}+\frac{\varepsilon_{\psi}%
}{2}e^{\kappa\phi}\dot{\psi}^{2}-V\left(  \phi\right)  ~, \label{sp.09}%
\end{equation}
while the effective parameter of state is given by the expression
\begin{equation}
w_{eff}=\frac{p_{eff}}{\rho_{eff}}=\frac{\varepsilon_{\phi}\dot{\phi}%
^{2}+\varepsilon_{\psi}e^{\kappa\phi}\dot{\psi}^{2}-2V\left(  \phi\right)
}{\varepsilon_{\phi}\dot{\phi}^{2}+\varepsilon_{\psi}e^{\kappa\phi}\dot{\psi
}^{2}+2V\left(  \phi\right)  }~. \label{sp.10}%
\end{equation}

We follow \cite{an2} and we rewrite the field equations (\ref{sp.04}),
(\ref{sp.07}) by using the variables \qquad%
\begin{equation}
\rho_{\phi}=\frac{\varepsilon_{\phi}}{2}\dot{\phi}^{2}+V\left(  \phi\right)
~,~p_{\phi}=\frac{\varepsilon_{\phi}}{2}\dot{\phi}^{2}-V\left(  \phi\right)
~, \label{sp.11}%
\end{equation}%
\begin{equation}
\rho_{\psi}=\frac{\varepsilon_{\psi}}{2}e^{\kappa\phi}\dot{\psi}^{2}%
~,~p_{\psi}=\frac{\varepsilon_{\psi}}{2}\varepsilon_{\psi}e^{\kappa\phi}%
\dot{\psi}^{2}~, \label{sp.12}%
\end{equation}
that is,
\begin{equation}
3H^{2}=\rho_{\phi}+\rho_{\psi}~, \label{sp.13}%
\end{equation}%
\begin{equation}
2\dot{H}+3H^{2}=-\left(  p_{\phi}+p_{\psi}\right)  ~, \label{sp.14}%
\end{equation}%
\begin{equation}
\dot{\rho}_{\phi}+3H\left(  \rho_{\phi}+p_{\phi}\right)  =\dot{\phi}%
\frac{\partial}{\partial\phi}p_{\psi}~ \label{sp.15}%
\end{equation}
and
\begin{equation}
\dot{\rho}_{\psi}+3H\left(  \rho_{\psi}+p_{\psi}\right)  =-\dot{\phi}%
\frac{\partial}{\partial\phi}p_{\psi}~. \label{sp.16}%
\end{equation}
from which it is clear how the two scalar fields interact. As we mentioned
above, the Chiral model can describe the whole dark sector of the universe,
that is, the dark matter and the dark energy, and there are various
observational results in the literature which support an interaction between
these two fluids \cite{in1,in2,in3,in4,in5,in6}.

\section{Asymptotic dynamics}

\label{sec3}

In order to continue with the study of dynamics for the field equations we
define the new dimensionless variables \cite{an2}
\begin{equation}
x=\frac{\dot{\phi}}{\sqrt{6}H}~,~y=\frac{\sqrt{V\left(  \phi\right)  }}%
{\sqrt{3}H}~,~z=\frac{e^{\frac{\kappa}{2}}\dot{\psi}}{\sqrt{6}}~,~\lambda
=\frac{V_{,\phi}}{V} \label{sp.17}%
\end{equation}
where the field equations (\ref{sp.04}), (\ref{sp.07}) are written in the
equivalent form of the algebraic-differential system%
\begin{equation}
\frac{dx}{d\ln a}=\frac{1}{2}\left(  3\varepsilon_{\phi}x^{3}-3x\left(
1+y^{2}-\varepsilon_{\psi}z^{2}\right)  +\frac{\sqrt{6}}{\varepsilon_{\phi}%
}\left(  \varepsilon_{\psi}\kappa z^{2}-\lambda y^{2}\right)  \right)  ~,
\label{sp.18}%
\end{equation}%
\begin{equation}
\frac{dy}{d\ln a}=\frac{1}{2}y\left(  3\left(  1-y^{2}+\varepsilon_{\phi}%
x^{2}+\varepsilon_{\psi}z^{2}\right)  +\sqrt{6}\lambda x\right)  ~,~
\label{sp.19}%
\end{equation}%
\begin{equation}
\frac{dz}{d\ln a}=\frac{1}{2}z\left(  3\left(  \varepsilon_{\phi}%
x^{2}-3-3y^{2}+\varepsilon_{\psi}z^{2}\right)  -\sqrt{6}\kappa~x\right)  ~
\label{sp.20}%
\end{equation}
and
\begin{equation}
\frac{d\lambda}{d\ln a}=\sqrt{6}x\lambda\left(  \Gamma\left(  \lambda\right)
-1\right)  ~,~\Gamma\left(  \lambda\right)  =\frac{V_{\phi\phi}V}{\left(
V_{\phi}\right)  ^{2}}~, \label{sp.21}%
\end{equation}
with algebraic constraint%
\begin{equation}
1-\varepsilon_{\phi}x^{2}-y^{2}-\varepsilon_{\psi}z^{2}=0. \label{sp.22}%
\end{equation}

Because of the algebraic constraint (\ref{sp.22}) the dynamical system
(\ref{sp.18})-(\ref{sp.21}) can be reduced by one dimension into a
three-dimensional system, while when $V\left(  \phi\right)  =V_{0}%
e^{\sigma\phi}$, $\sigma=const.,~$it follows $\Gamma\left(  \lambda\right)
=1$, which means that $\lambda=const$. Thus the final dynamical system has
dimension two.

Furthermore, the parameter for the equation of state for the effective fluid
is calculated to be%
\begin{equation}
w_{eff}\left(  x,y,z\right)  =\varepsilon_{\phi}x^{2}-y^{2}+\varepsilon_{\psi
}z^{2}. \label{sp.23}%
\end{equation}

What is important is to find the range of the variables $\left(  x,y,z\right)
$. For the model A with $\left(  \varepsilon_{\phi},\varepsilon_{\psi}\right)
=\left(  +1,+1\right)  $, variables $\left(  x,y,z\right)  $ are defined on
the surface of the three-dimensional unitary sphere $S^{3}$. This is not true
for the remainder of the models which are: model B with $\left(
\varepsilon_{\phi},\varepsilon_{\psi}\right)  =\left(  +1,-1\right)
,~$\ model C with $\left(  \varepsilon_{\phi},\varepsilon_{\psi}\right)
=\left(  -1,+1\right)  $ and model D where now $\left(  \varepsilon_{\phi
},\varepsilon_{\psi}\right)  =\left(  -1,-1\right)  $. In these three models,
the variables $\left(  x,y,z\right)  $ are not bounded.

\section{Stationary points for Exponential potential}

\label{sec4}

For the exponential potential, $V\left(  \phi\right)  =V_{0}e^{\sigma\phi}$,
the dynamical system, (\ref{sp.18})-(\ref{sp.21}), reduces to a
two-dimensional system, where $y^{2}=1-\varepsilon_{\phi}x^{2}-\varepsilon
_{\psi}z^{2}$. The stationary points are found to be%
\[
P_{1}^{\left(  \pm\right)  }=\left(  \pm\frac{1}{\sqrt{\varepsilon_{\phi}}%
},0,0\right)  ~,~P_{2}=\left(  -\frac{\lambda}{\varepsilon_{\phi}\sqrt{6}%
},\sqrt{1-\frac{\lambda^{2}}{6\varepsilon_{\phi}}},0\right)  ~
\]
and%
\[
P_{3}^{\left(  \pm\right)  }=\left(  -\frac{\sqrt{6}}{\kappa+\lambda}%
,\sqrt{\frac{\kappa}{\kappa+\lambda}},\pm\sqrt{\frac{\lambda\left(
\kappa+\lambda\right)  -6\varepsilon_{\phi}}{\varepsilon_{\psi}\left(
\kappa+\lambda\right)  ^{2}}}\right)  ~\text{.}%
\]
At the stationary points $P_{1}^{\left(  \pm\right)  },~P_{2}$ only the scalar
field $\phi$ contributes in the cosmological fluid, while the second field
appears at the stationary points $P_{3}^{\left(  \pm\right)  }$.

We continue by studying in detail, the viability of the stationary points and
their stability for the four models of our analysis.

\subsection{Model A~$\left(  \varepsilon_{\phi},\varepsilon_{\psi}\right)
=\left(  +1,+1\right)  $}

For the Chiral model with ~$\left(  \varepsilon_{\phi},\varepsilon_{\psi
}\right)  =\left(  +1,+1\right)  $ and for the exponential potential the
asymptotic dynamics have been studied before in \cite{an3}. For a completeness
of our study we reproduce these results. In this model the stationary points
are $P_{1}^{\left(  \pm\right)  }\left(  A\right)  =\left(  \pm1,0,0\right)
~,~P_{2}\left(  A\right)  =\left(  -\frac{\lambda}{\sqrt{6}},\sqrt
{1-\frac{\lambda^{2}}{6}},0\right)  $ and $P_{3}^{\left(  \pm\right)  }\left(
A\right)  =\left(  -\frac{\sqrt{6}}{\kappa+\lambda},\sqrt{\frac{\kappa}%
{\kappa+\lambda}},\pm\sqrt{\frac{\lambda\left(  \kappa+\lambda\right)
-6}{\left(  \kappa+\lambda\right)  ^{2}}}\right)  $. \begin{figure}[ptb]
\centering\includegraphics[width=0.45\textwidth]{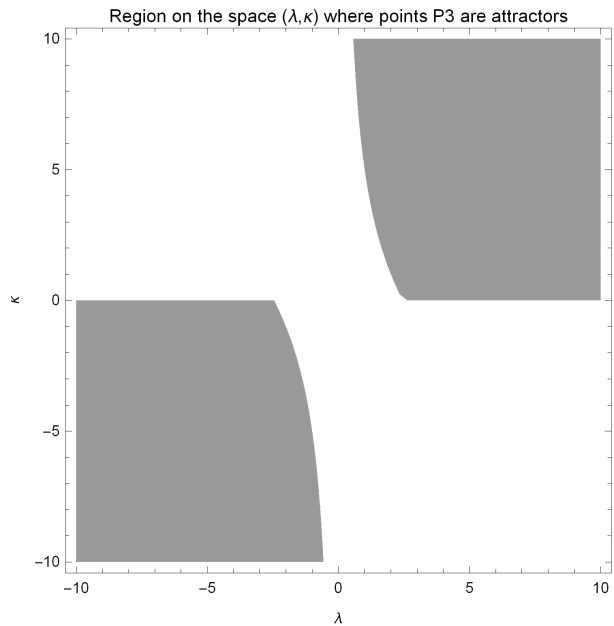} \caption{Region
plots in the space of variables $\left(  \lambda,\kappa\right)  $ where the
stationary points $P_{3}^{\left(  \pm\right)  }\left(  A\right)  ~$are
attractors.}%
\label{fig1A}%
\end{figure}

Points $P_{1}^{\left(  \pm\right)  }\left(  A\right)  $ are always physically
acceptable and describe universes in which only the kinetic part of the scalar
field $\phi$ contributes in the total cosmological fluid. Indeed
$w_{eff}\left(  P_{1}^{\left(  \pm\right)  }\left(  A\right)  \right)  =1$.
This means that the stationary points describe stiff fluid solutions. The
eigenvalues of the linearized system are $e_{1}\left(  P_{1}^{\left(
\pm\right)  }\left(  A\right)  \right)  =6\pm\sqrt{6}\lambda~,~e_{2}\left(
P_{1}^{\left(  \pm\right)  }\left(  A\right)  \right)  =\mp\sqrt{\frac{3}{2}%
}\kappa$, from which we infer that $P_{1}^{\left(  +\right)  }\left(
A\right)  $ is an attractor for$\lambda<-\sqrt{6}$ and $\kappa>0$ while
$P_{1}^{\left(  -\right)  }\left(  A\right)  $ is an attractor when
$\lambda>\sqrt{6}$ and $\kappa<0$.

\begin{figure}[ptb]
\centering\includegraphics[width=0.8\textwidth]{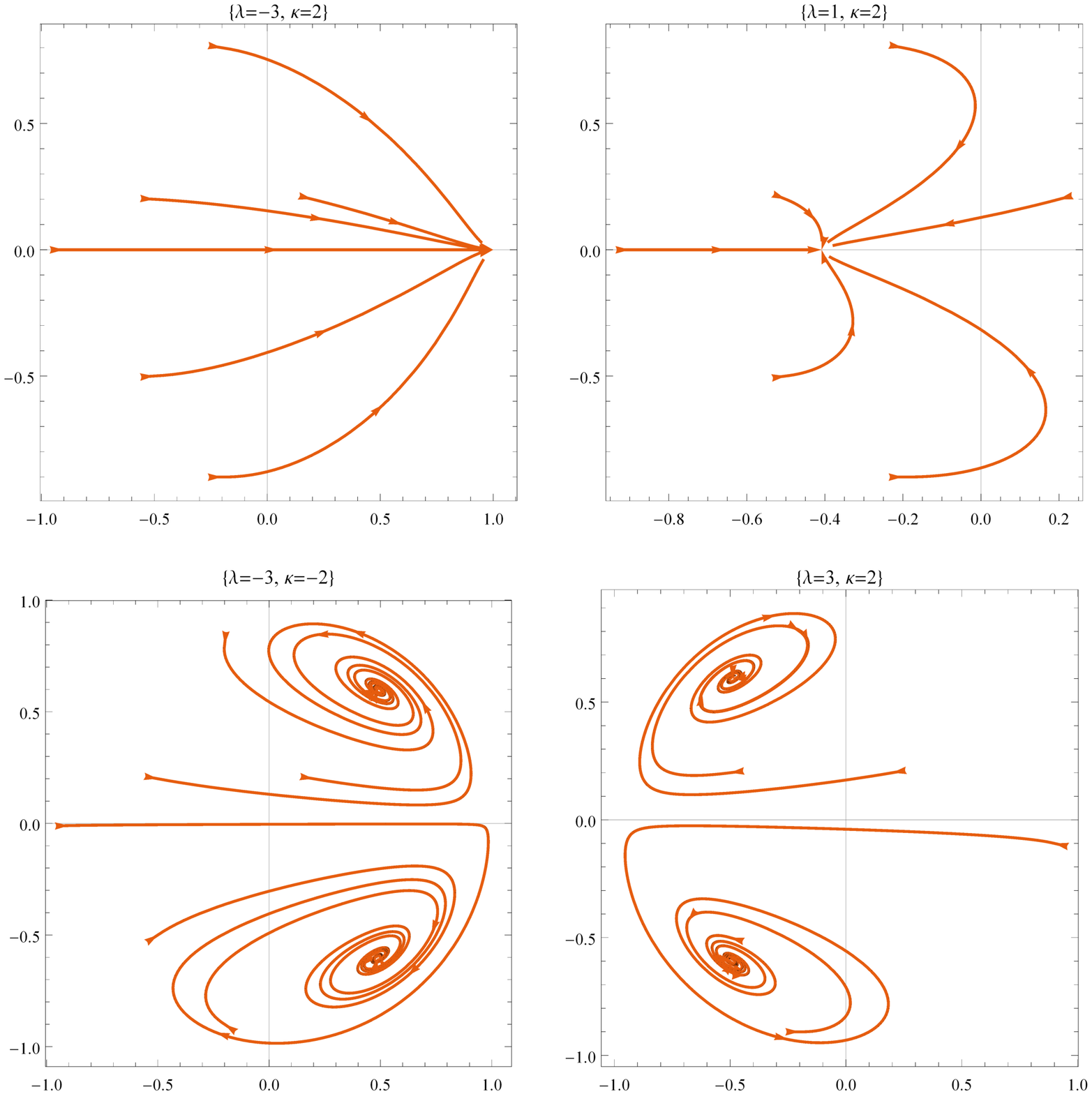}\caption{Phase space
portrait in the space of variables $\left(  x,z\right)  $ for the dynamical
system (\ref{sp.18}), (\ref{sp.20}) and for $\left(  \varepsilon_{\phi
},\varepsilon_{\psi}\right)  =\left(  +1,+1\right)  $. The figures are for
different values of the free parameters $\left\{  \lambda,\kappa\right\}  $.
In the plots of the first row the unique attractor is point $P_{2}$, while in
the second row the attractors are points $P_{3}^{\left(  \pm\right)  }$. }
\label{fig2}
\end{figure}

The stationary point $P_{2}\left(  A\right)  $ is physically acceptable for
$\left\vert \lambda\right\vert <\sqrt{6}$. The point describes a scaling
solution $w_{eff}\left(  P_{2}\left(  A\right)  \right)  =-1+\frac{\lambda
^{2}}{3}$, from which it follows that $w_{eff}\left(  P_{2}\left(  A\right)
\right)  <-\frac{1}{3}$ if and only if $\left\vert \lambda\right\vert
<\sqrt{2}$. The eigenvalues of the linearized system are determined to be
$e_{1}\left(  P_{2}\left(  A\right)  \right)  =\frac{\left(  \lambda
^{2}-6\right)  }{2}~,~e_{2}\left(  P_{2}\left(  A\right)  \right)  =\frac
{1}{2}\left(  \lambda^{2}+\kappa\lambda-6\right)  $. Thus the stationary point
is an attractor when~$\left\{  -\sqrt{6}<\lambda<0,~\kappa>\frac{6-\lambda
^{2}}{\lambda}\right\}  ~\cup~\left\{  0<\lambda<\sqrt{6},\kappa
<\frac{6-\lambda^{2}}{\lambda}\right\}  \cup\left\{  \lambda=0\right\}  $.

Stationary points $P_{3}^{\left(  \pm\right)  }\left(  A\right)  $ are real
and physically acceptable when $\left\{  \lambda,\kappa\right\}  $ are
constrained as $\left\{  \lambda\leq-\sqrt{6},\kappa<0\right\}  \cup\left\{
-\sqrt{6}<\lambda<0,\kappa<\frac{6-\lambda^{2}}{\lambda}\right\}  $ and
$\left\{  0<\lambda<\sqrt{6},\kappa>\frac{6-\lambda^{2}}{\lambda}\right\}
\cup\left\{  \lambda\geq\sqrt{6},\kappa>0\right\}  $. The effective fluid has
equation of state $w_{eff}\left(  P_{3}^{\left(  \pm\right)  }\left(
A\right)  \right)  =1-\frac{2\kappa}{\kappa+\lambda}$, from which we infer
that the exact solutions at the points describe accelerated universes when
$\left\{  \lambda\leq-\sqrt{2},\kappa<2\lambda\right\}  \cup\left\{  -\sqrt
{2}<\lambda<0,\kappa<\frac{6-\lambda^{2}}{\lambda}\right\}  $ and $\left\{
0<\lambda<\sqrt{2},\kappa>\frac{6-\lambda^{2}}{\lambda}\right\}  \cup\left\{
\lambda\geq\sqrt{2},\kappa>2\lambda\right\}  $. The eigenvalues of the
linearized system around the stationary points $P_{3}^{\left(  \pm\right)
}\left(  A\right)  $ are derived as $e_{\pm}\left(  P_{3}^{\left(  \pm\right)
}\left(  A\right)  \right)  =-\frac{3\kappa}{2\left(  \kappa+\lambda\right)
}\pm\frac{i\sqrt{3\kappa}}{2\left(  \kappa+\lambda\right)  }\sqrt{\Delta}$,
where $\Delta=4\lambda^{3}+8\kappa\lambda^{2}+\lambda\left(  \kappa
^{2}-6\right)  -27\kappa$. The real part of eigenvalue, $e_{-}\left(
P_{3}^{\left(  \pm\right)  }\left(  A\right)  \right) , $ is always negative
when $\frac{\kappa}{\left(  \kappa+\lambda\right)  }>0$. Thus in this case the
stationary points can be saddles or attractors. The region plot in the space
of the variables $\left(  \lambda,\kappa\right)  $, in which the exact
solutions at the points $P_{3}^{\left(  \pm\right)  }\left(  A\right)  $ are
stable, is presented in Fig. \ref{fig1A}.

In Fig. \ref{fig2} there are the phase space portraits in the space of
variables $\left(  x,z\right)  $ for the dynamical system (\ref{sp.18}),
(\ref{sp.20}). The figures are for different values of the free parameters
$\left\{  \lambda,\kappa\right\}  $ so that to have different stationary
points as attractors.

\subsection{Model B~$\left(  \varepsilon_{\phi},\varepsilon_{\psi}\right)
=\left(  +1,-1\right)  $}

For the model with $\left(  \varepsilon_{\phi},\varepsilon_{\psi}\right)
=\left(  +1,-1\right)  $ the admitted stationary points of the gravitational
field equations are $P_{1}^{\left(  \pm\right)  }\left(  B\right)  =\left(
\pm1,0,0\right)  ,~P_{2}\left(  B\right)  =\left(  -\frac{\lambda}{\sqrt{6}%
},\sqrt{1-\frac{\lambda^{2}}{6}},0\right)  $ and $P_{3}^{\left(  \pm\right)
}\left(  B\right)  =\left(  -\frac{\sqrt{6}}{\kappa+\lambda},\sqrt
{\frac{\kappa}{\kappa+\lambda}},\pm\sqrt{\frac{6-\lambda\left(  \kappa
+\lambda\right)  }{\left(  \kappa+\lambda\right)  ^{2}}}\right)  $.

For the stationary Points $P_{1}^{\left(  \pm\right)  }\left(  B\right)  $ and
$P_{2}\left(  B\right)  $ we find the same physical properties as with the
corresponding points of model A. Furthermore, points $P_{3}^{\left(
\pm\right)  }\left(  B\right)  $ are real and physically acceptable when
$\left\{  \lambda\leq-\sqrt{6}~,~\kappa>-\lambda\right\}  \cup\left\{
-\sqrt{6}<\lambda<0,~\kappa>-\lambda\text{ or }\frac{6-\lambda^{2}}{\lambda
}<\kappa<0\right\}  ~$\ and $\left\{  0<\lambda<\sqrt{6},~\kappa
<-\lambda\text{ or }0<\kappa<\frac{6-\lambda^{2}}{\lambda}\right\}
\cup\left\{  \lambda\subset\sqrt{6}~,~\kappa<-\lambda\right\}  $. The
eigenvalues of the linearized system near the points $P_{3}^{\left(
\pm\right)  }\left(  B\right)  $ have the same functional form as points
$P_{3}^{\left(  \pm\right)  }\left(  A\right)  $. Thus for this model the
exact solutions at the stationary points are always unstable, while the
stationary points are always saddle points.

\begin{figure}[ptb]
\centering\includegraphics[width=0.8\textwidth]{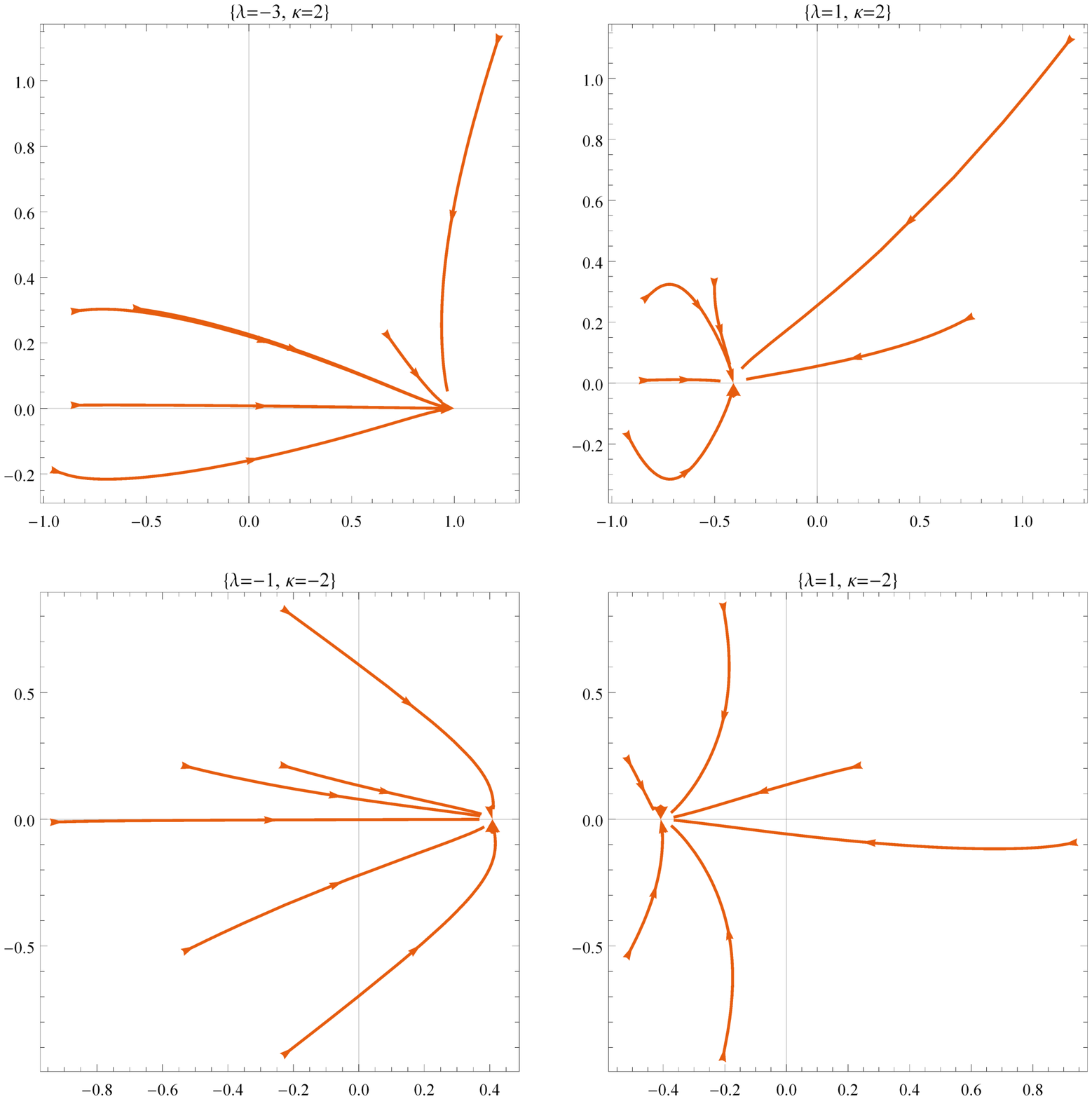} \caption{Phase
space portrait in the space of variables $\left(  x,z\right)  $ for the
dynamical system (\ref{sp.18}), (\ref{sp.20}) and for $\left(  \varepsilon
_{\phi},\varepsilon_{\psi}\right)  =\left(  +1,-1\right)  $. The figures are
for different values of the free parameters $\left\{  \lambda,\kappa\right\}
$. We observe that the unique attractor is point $P_{2}\left(  B\right)  $. }%
\label{fig3}%
\end{figure}

Phase space portraits for the variables $\left(  x,z\right)  $ of the
gravitational field equations for $\left(  \varepsilon_{\phi},\varepsilon
_{\psi}\right)  =\left(  +1,-1\right)  $ are presented in Fig. \ref{fig3},
from where we observe that the unique attractor is point $P_{2}\left(
B\right)  $. \begin{figure}[ptb]
\centering\includegraphics[width=0.8\textwidth]{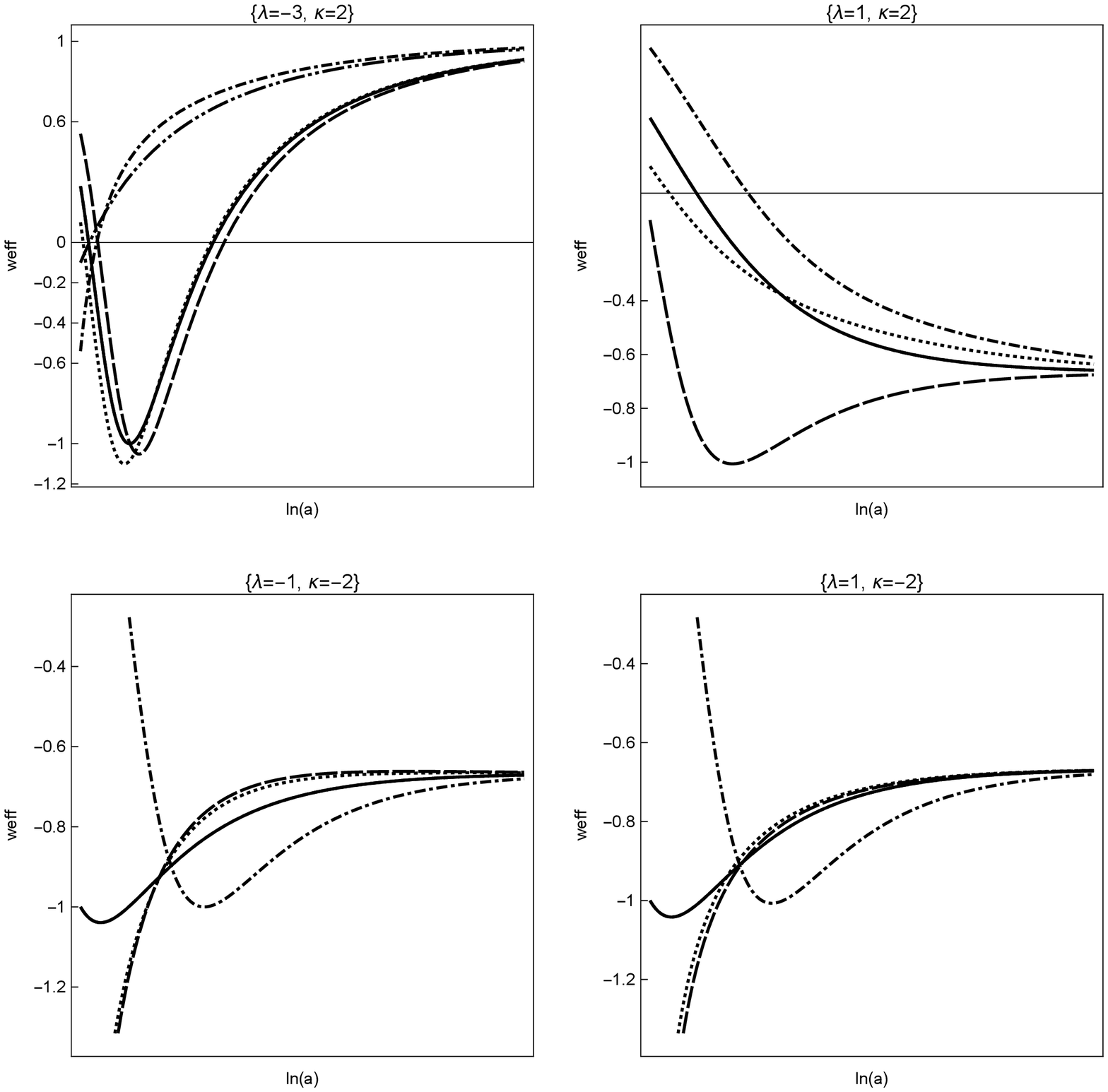}
\caption{Qualitative evolution of $w_{eff}$ for the numerical solutions of the
phase portraits in Fig. \ref{fig3}}%
\label{fig4A}%
\end{figure}

In Fig. \ref{fig4A} we present the qualitative evolution for the parameter for
the equation of state for the effective fluid, \ $w_{eff}$, for the solutions
presented in the phase space of Fig. \ref{fig3}. It is clear that the equation
of state parameter can cross the phantom divide line twice. That is it can
start from the $w_{eff}>-1$ then take values $w_{eff}<-1$ and cross again the
limit and end with $w_{eff}>-1$.

It is important to mention here that in contrast to model A, the variables
$\left(  x,z\right)  $ are not constrained and they can take values at .

\subsubsection{Analysis at infinity}

\begin{figure}[ptb]
\centering\includegraphics[width=0.8\textwidth]{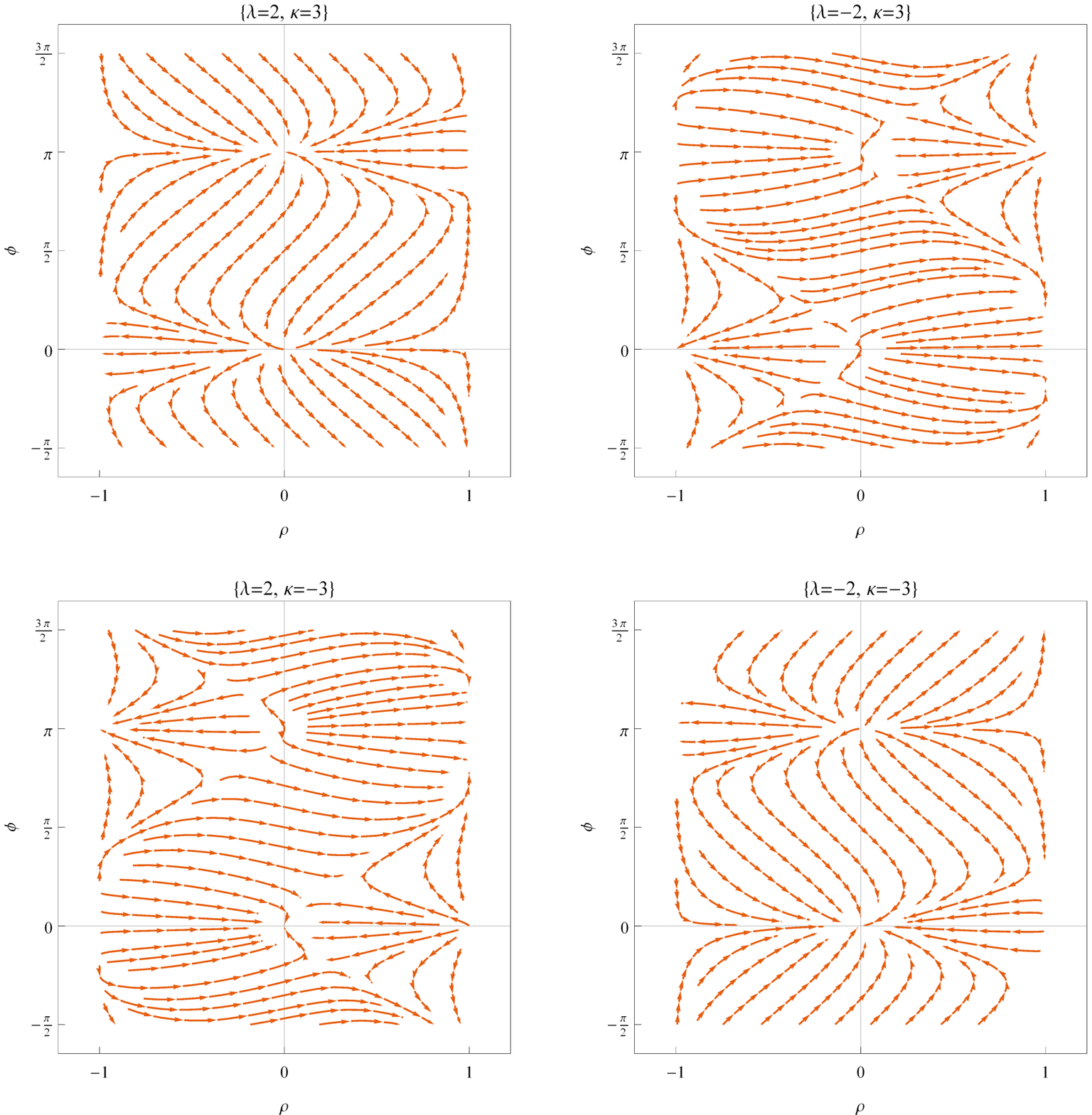} \caption{Phase
space portrait in the space of variables $\left(  \rho,v\right)  $ for the
dynamical system (\ref{s1}), (\ref{s2}). The figures are for different values
of the free parameters $\left\{  \lambda,\kappa\right\}  $.}%
\label{fig5}%
\end{figure}

In order to perform the analysis at infinity we define the new variables
\begin{equation}
x=\frac{\cos u\cos v}{\rho}~,~y=\frac{\cos u\sin v}{\rho}~,~z=\frac{\sin
u}{\rho}~,
\end{equation}
where when $\rho\rightarrow0$, parameters $\left(  x,y,z\right)  $ have values
at infinity. The constraint (\ref{sp.22}) becomes $\rho^{2}-\cos\left(
2\theta\right)  =0$, while the field equations reads%
\begin{equation}
\rho^{\prime}=\frac{1}{2}\left(  \sqrt{6\left(  1-\rho^{4}\right)  }\kappa
\rho\cos v\sin\left(  \frac{1}{2}\arccos\rho^{2}\right)  -3\sin^{2}v\left(
\rho^{4}-1\right)  \right)  ~, \label{s1}%
\end{equation}%
\begin{equation}
v^{\prime}=\frac{1}{4}\left(  12\rho\cos v+\sqrt{6}\left(  \kappa+\lambda
+\rho^{2}\left(  \lambda-\kappa\right)  \right)  \frac{\sin v}{\cos\left(
\frac{1}{2}\arccos\rho^{2}\right)  }\right)  ~, \label{s2}%
\end{equation}
where $\rho^{\prime}=\rho\frac{d\rho}{d\ln a}$. The stationary points
$Q=\left(  \rho,v\right)  ~$with $\rho=0$ are those with $v_{1}=0$ and
$v_{2}=\pi$. The points with coordinates~$Q_{1}=\left(  0,0\right)  $\ and
$Q_{2}=\left(  0,\pi\right)  $ provide physical solutions where only the
kinetic parts of the scalar field contributes, that is, the physical solution
describes a stiff fluid.

As far as the stability of the stationary points is concerned, it follows that
$Q_{1}$ is an attractor when $\left\{  \lambda<0,~\kappa<0\right\}  $ or
$\left\{  \lambda>0,~\kappa<-\lambda\right\}  $, while $Q_{2}$ is an attractor
when~$\left\{  \lambda<0,~\kappa>-\lambda\right\}  $ and $\left\{
\lambda>0,\kappa>0\right\}  $. In Fig. \ref{fig5} we present the phase space
portrait for the dynamical system (\ref{s1}), (\ref{s2}).

\subsection{Model C~$\left(  \varepsilon_{\phi},\varepsilon_{\psi}\right)
=\left(  -1,+1\right)  $}

In the case where $\phi$ is a phantom field, that is, $\left(  \varepsilon
_{\phi},\varepsilon_{\psi}\right)  =\left(  -1,+1\right)  $, the real
stationary points are the $P_{2}\left(  C\right)  =\left(  \frac{\lambda
}{\sqrt{6}},\sqrt{1+\frac{\lambda^{2}}{6}},0\right)  $ and $P_{3}^{\left(
\pm\right)  }\left(  C\right)  =\left(  -\frac{\sqrt{6}}{\kappa+\lambda}%
,\sqrt{\frac{\kappa}{\kappa+\lambda}},\pm\sqrt{\frac{\lambda\left(
\kappa+\lambda\right)  +6}{\left(  \kappa+\lambda\right)  ^{2}}}\right)  $.
The exact solution at point $P_{2}\left(  C\right)  $ describes a universe
where the effective fluid has an equation of state parameter $w_{eff}%
=-1-\frac{\lambda^{2}}{3}$, which means that $w_{eff}<-1$, crosses the phantom
divide line. On the other hand, points $P_{3}^{\left(  \pm\right)  }\left(
C\right)  $ have the same physical properties with points $P_{3}^{\left(
\pm\right)  }\left(  A\right)  $, while points are real when $\lambda\left(
\kappa+\lambda\right)  +6>0$, that is, $\left\{  \lambda<0,\kappa
<-\frac{6+\lambda^{2}}{\lambda}\right\}  $ $\cup$ $\left\{  \lambda
>0,\kappa>-\frac{6+\lambda^{2}}{\lambda}\right\}  \cup\left\{  \kappa\in%
\mathbb{R}
,\lambda=0\right\}  $.

As far as the stability is concerned, the eigenvalues of the linearized system
around point $P_{2}\left(  C\right)  $ are $e_{1}\left(  P_{2}\left(
C\right)  \right)  =-3-\frac{\lambda^{2}}{2}~,~e_{2}\left(  P_{2}\left(
C\right)  \right)  =-\frac{1}{2}\lambda\left(  \kappa+\lambda\right)  +6,$
from where we infer that the point is an attractor when $\left\{
\lambda<0,\kappa<-\frac{6+\lambda^{2}}{\lambda}\right\}  $ $\cup$ $\left\{
\lambda>0,\kappa>-\frac{6+\lambda^{2}}{\lambda}\right\}  \cup\left\{
\kappa\in%
\mathbb{R}
,\lambda=0\right\}  $. Moreover, for points $P_{3}^{\left(  \pm\right)
}\left(  C\right)  $ we calculate the eigenvalues~$e_{1}\left(  P_{3}^{\left(
\pm\right)  }\left(  C\right)  \right)  =-\frac{3\kappa}{2\left(
\kappa+\lambda\right)  }\pm\frac{\sqrt{3\kappa}}{2\left(  \kappa
+\lambda\right)  }\sqrt{\bar{\Delta}},~$with~$\bar{\Delta}=4\lambda
^{3}+8\kappa\lambda^{2}+\lambda\left(  \kappa^{2}+6\right)  +27\kappa$, from
where it follows that the exact solutions at the points are unstable, while
points are saddles points for $\left\{  \lambda<0,~-\lambda<\kappa
<-\frac{6+\lambda^{2}}{\lambda}\right\}  $ $\cup$ $\left\{  \lambda
>0,-\frac{6+\lambda^{2}}{\lambda}<\kappa<-\lambda\right\}  \ $and $\left\{
\kappa>0,\lambda\geq0~\right\}  \cup\left\{  \kappa<0,\lambda\leq0\right\}  $,
otherwise the points are sources.

\begin{figure}[ptb]
\centering\includegraphics[width=0.8\textwidth]{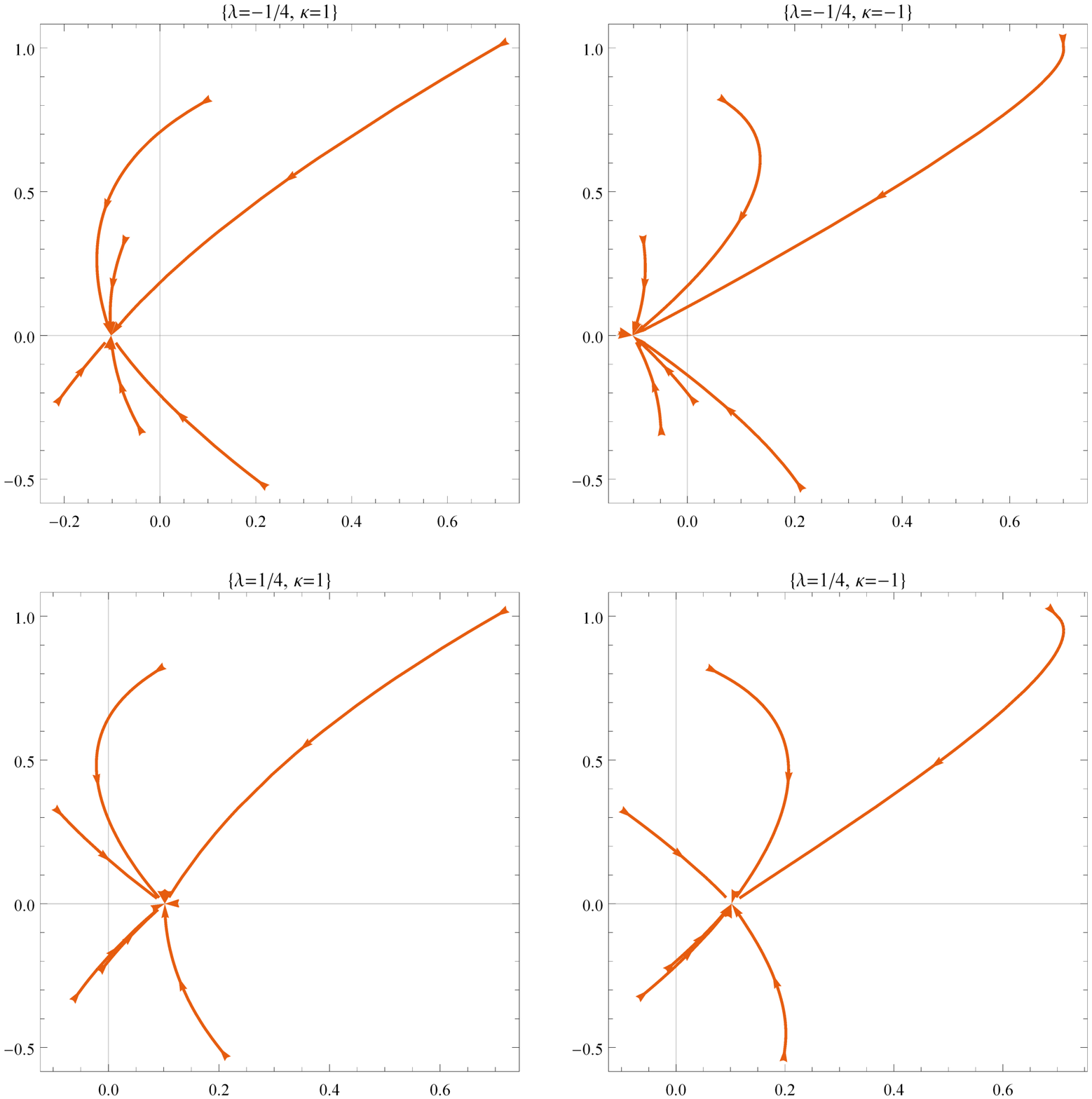} \caption{Phase
space portrait in the space of variables $\left(  x,z\right)  $ for the
dynamical system (\ref{sp.18}), (\ref{sp.20}) and for $\left(  \varepsilon
_{\phi},\varepsilon_{\psi}\right)  =\left(  -1,+1\right)  $. The figures are
for different values of the free parameters $\left\{  \lambda,\kappa\right\}
$. We observe that the unique attractor is point $P_{2}\left(  B\right)  $. }%
\label{fig6}%
\end{figure}

\begin{figure}[ptb]
\centering\includegraphics[width=0.8\textwidth]{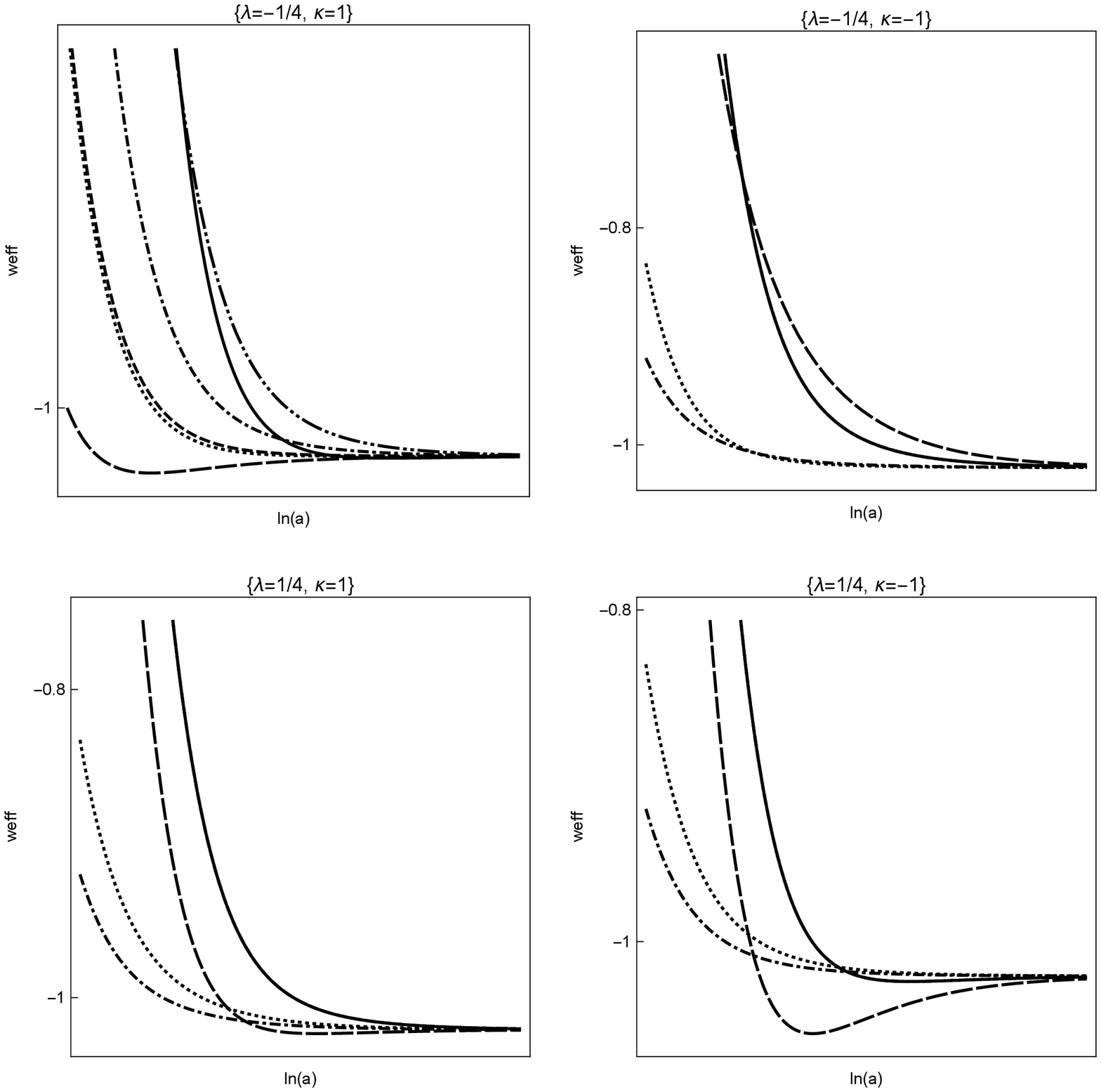}
\caption{Qualitative evolution of $w_{eff}$ for the numerical solutions of the
phase portraits in Fig. \ref{fig6}}%
\label{fig7}%
\end{figure}

Phase space portraits of the dynamical system and the qualitative behaviour of
the effective equation of state parameter $w_{eff}$ are presented in Figs.
\ref{fig6} and \ref{fig7}. It is obvious that $w_{eff}$ crosses the phantom
divide line.

\subsubsection{Analysis at infinity}

As in the model B, parameters can take values at infinity. Thus we define the
new coordinates%
\begin{equation}
x=\frac{\cos u\cosh v}{\rho}~,~y=\frac{\cos u\sinh v}{\rho}~,~z=\frac{\sin
u}{\rho}~,
\end{equation}
where the dynamical system becomes%
\begin{equation}
\rho^{\prime}=\frac{1}{2}\left(  \rho^{2}+1\right)  \left(  \kappa\rho
\sqrt{3\left(  1-\rho^{2}\right)  }\cosh v-3\left(  \rho^{2}-1\right)
\sinh^{2}v\right)  ~, \label{s3}%
\end{equation}%
\begin{equation}
v^{\prime}=\frac{1}{4}\left(  \sqrt{6}\left(  \kappa+\lambda+\left(
\kappa-\lambda\right)  \rho^{2}\right)  \frac{\sinh v}{\cos\left(  \frac{1}%
{2}\arccos\left(  -\frac{\rho^{2}}{2}\right)  \right)  }+6\rho\sinh\left(
2v\right)  \right)  ~, \label{s4}%
\end{equation}
where $\rho^{2}+\cos\left(  2\theta\right)  =0$. We find only one stationary
point at the infinity with coordinates $\left(  \rho,v\right)  $ to be
$Q_{3}=\left(  0,0\right)  $. The point describes a solution in which only the
kinetic parts of the scalar fields contribute to the cosmological fluid, while
the eigenvalues of the linearized system are determined to be $e_{1}\left(
Q_{3}\right)  =\frac{\sqrt{3}}{2}\kappa,~e_{2}\left(  Q_{3}\right)
=\frac{\sqrt{3}}{2}\left(  \kappa+\lambda\right)  $, from where we infer that
the point is an attractor when $\kappa<0$ and $\kappa<-\lambda$.

\begin{figure}[ptb]
\centering\includegraphics[width=0.8\textwidth]{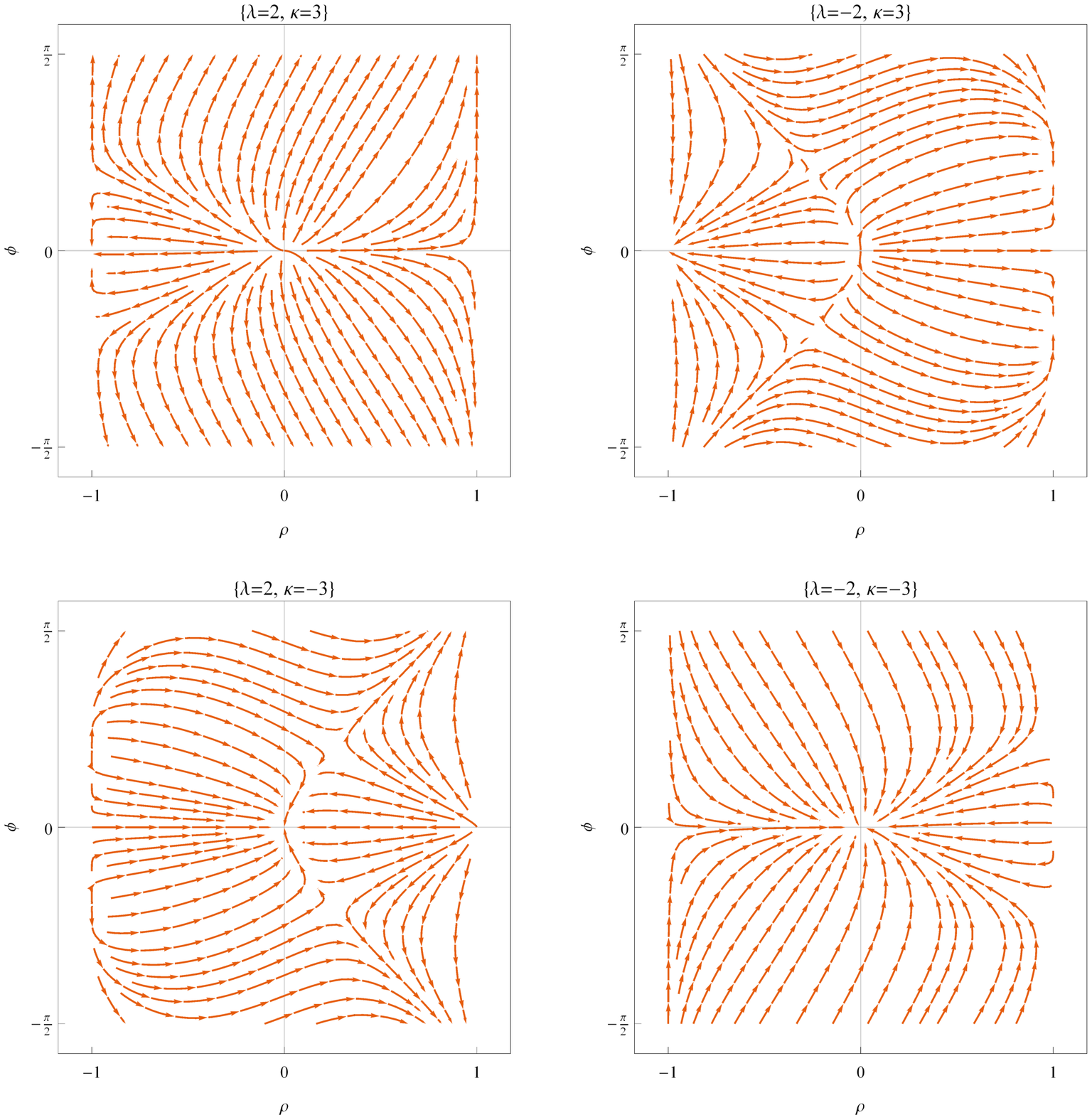} \caption{Phase
space portrait in the space of variables $\left(  \rho,v\right)  $ for the
dynamical system (\ref{s3}), (\ref{s4}). The figures are for different values
of the free parameters $\left\{  \lambda,\kappa\right\}  $.}%
\label{fig8}%
\end{figure}

The phase portrait of the dynamical system (\ref{s3}), (\ref{s4}) is presented
in Fig. \ref{fig8}. 

\subsection{Model D~$\left(  \varepsilon_{\phi},\varepsilon_{\psi}\right)
=\left(  -1,-1\right)  $}

\begin{figure}[ptb]
\centering\includegraphics[width=0.45\textwidth]{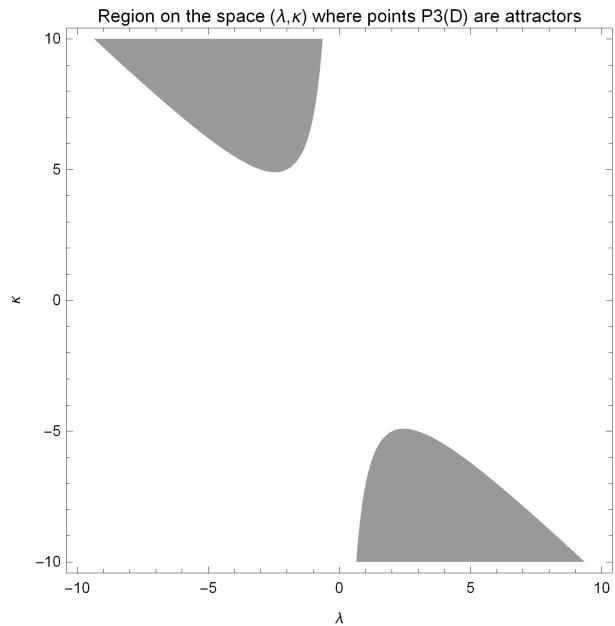} \caption{Region
plots in the space of variables $\left(  \lambda,\kappa\right)  $ for which
the stationary points $P_{3}^{\left(  \pm\right)  }\left(  D\right)  ~$are
attractors.}%
\label{fig9}%
\end{figure}

\begin{figure}[ptb]
\centering\includegraphics[width=0.8\textwidth]{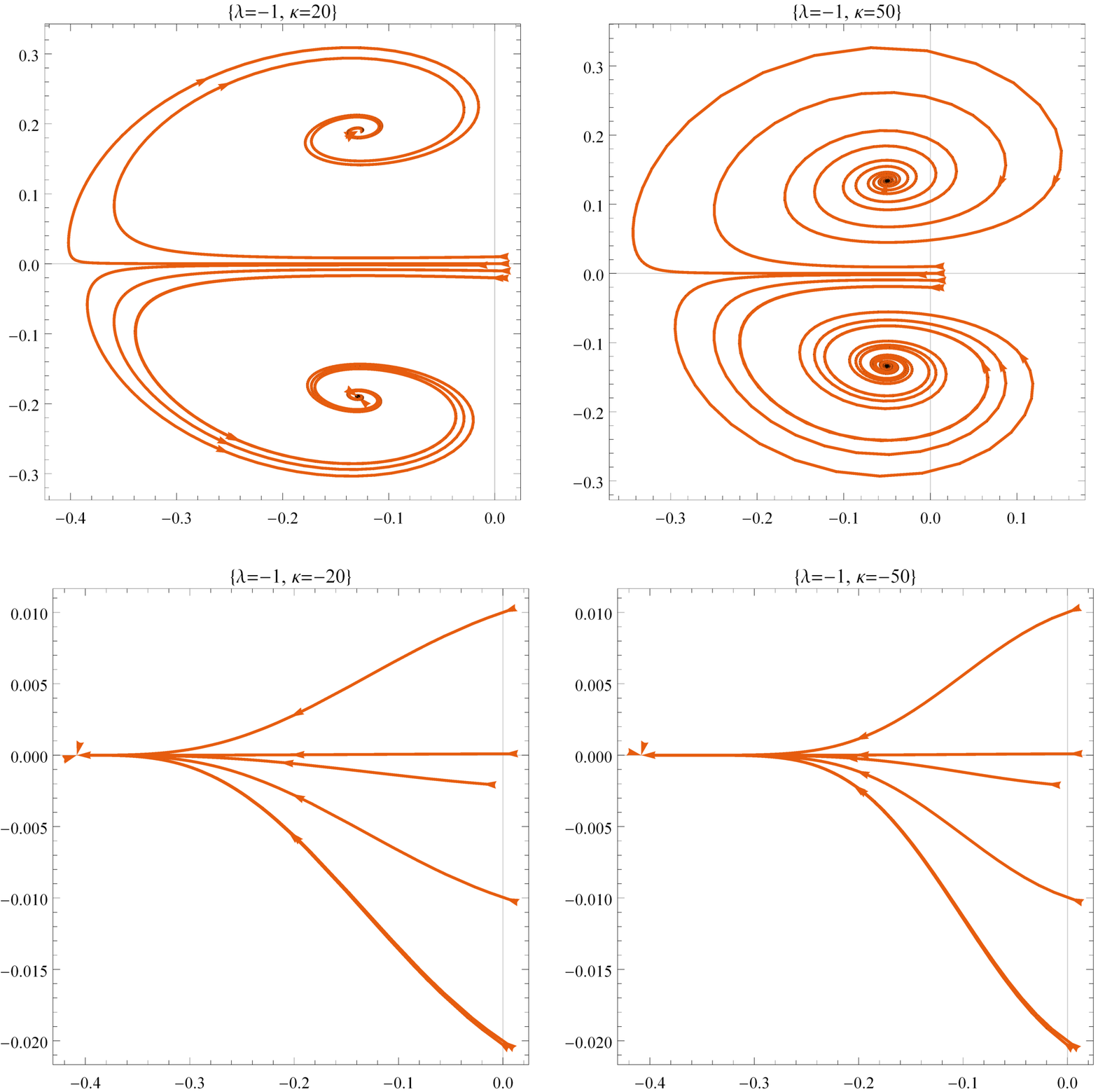} \caption{Phase
space portrait in the space of variables $\left(  x,z\right)  $ for the
dynamical system (\ref{sp.18}), (\ref{sp.20}) and for $\left(  \varepsilon
_{\phi},\varepsilon_{\psi}\right)  =\left(  -1,-1\right)  $. The figures are
for different values of the free parameters $\left\{  \lambda,\kappa\right\}
$.}%
\label{fig10}%
\end{figure}

\begin{figure}[ptb]
\centering\includegraphics[width=0.8\textwidth]{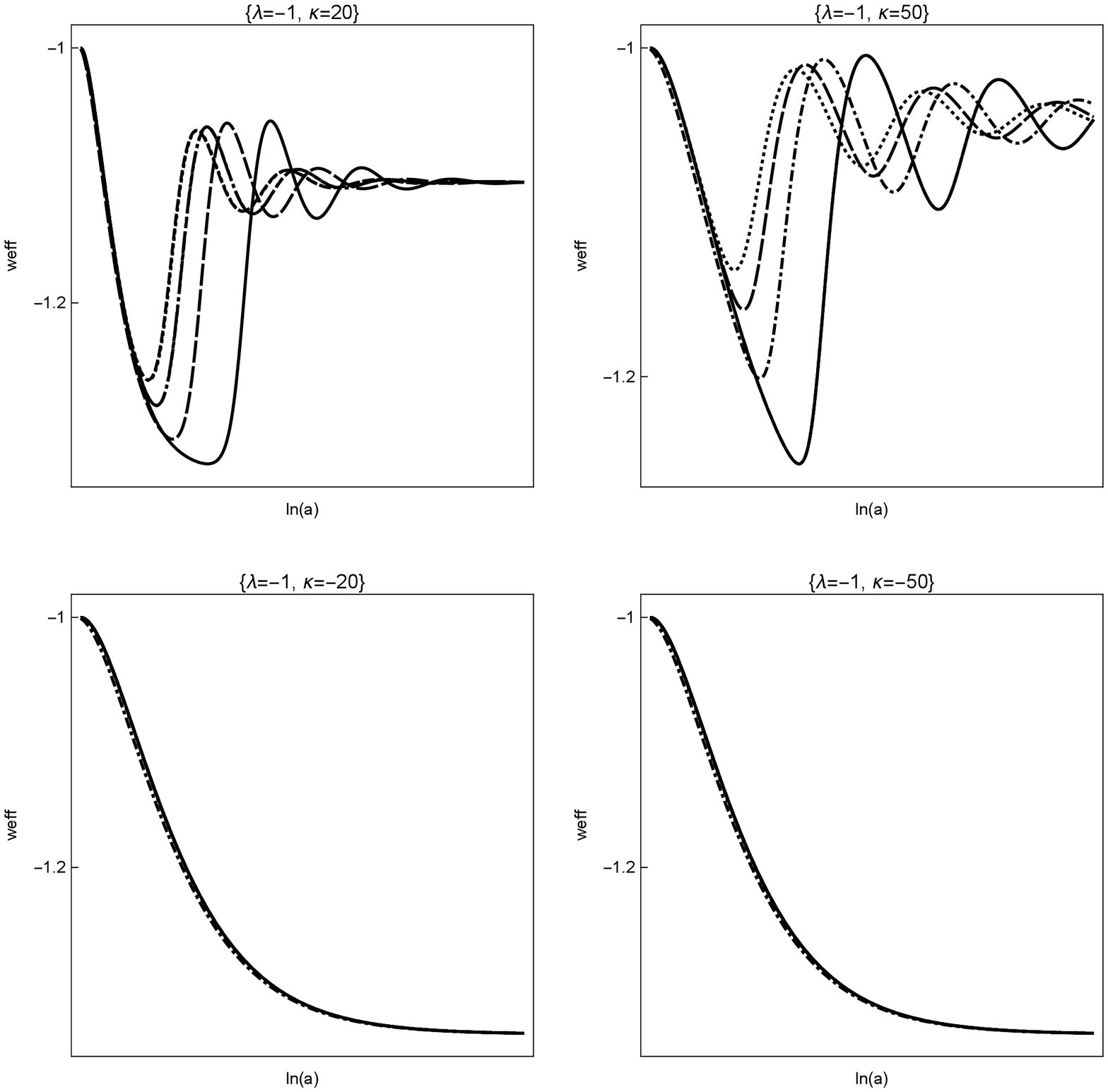}
\caption{Qualitative evolution of $w_{eff}$ for the numerical solutions of the
phase portraits in Fig. \ref{fig10}}%
\label{fig11}%
\end{figure}

For the last case of our interest, where $\left(  \varepsilon_{\phi
},\varepsilon_{\psi}\right)  =\left(  -1,-1\right)  $ the real stationary
points are the $P_{2}\left(  D\right)  =\left(  \frac{\lambda}{\sqrt{6}}%
,\sqrt{1+\frac{\lambda^{2}}{6}},0\right)  $ and $P_{3}^{\left(  \pm\right)
}\left(  D\right)  =\left(  -\frac{\sqrt{6}}{\kappa+\lambda},\sqrt
{\frac{\kappa}{\kappa+\lambda}},\pm\sqrt{-\frac{\lambda\left(  \kappa
+\lambda\right)  +6}{\left(  \kappa+\lambda\right)  ^{2}}}\right)  $. Point
$P_{2}\left(  D\right)  $ has same physical properties as $P_{2}\left(
C\right)  $, while points $P_{3}^{\left(  \pm\right)  }\left(  D\right)  $
have same physics as $P_{3}^{\left(  \pm\right)  }\left(  C\right)  $. Thus
the stability properties change. Indeed the eigenvalues of the linearized
system around the point $P_{2}\left(  D\right)  $ \ are $e_{1}\left(
P_{2}\left(  D\right)  \right)  =-3-\frac{\lambda^{2}}{2}~,~e_{2}\left(
P_{2}\left(  D\right)  \right)  =\frac{1}{2}\left(  -6-\lambda\left(
\kappa+\lambda\right)  \right)  $ from which we infer that $P_{2}\left(
D\right)  $ is an attractor when $\left\{  \lambda<0,\kappa<-\frac
{6+\lambda^{2}}{\lambda}\right\}  $ $\cup$ $\left\{  \lambda>0,\kappa
>-\frac{6+\lambda^{2}}{\lambda}\right\}  \cup\left\{  \kappa\in%
\mathbb{R}
,\lambda=0\right\}  $. Stationary points $P_{3}^{\left(  \pm\right)  }\left(
D\right)  $ are physically acceptable when $\lambda\left(  \kappa
+\lambda\right)  +6<0$, while the eigenvalues of the linearized system are
determined to be $e_{\pm}\left(  P_{3}^{\left(  \pm\right)  }\left(  D\right)
\right)  =-\frac{3\kappa}{2\left(  \kappa+\lambda\right)  }\pm\frac
{\sqrt{3\kappa}}{2\left(  \kappa+\lambda\right)  }\sqrt{\bar{\Delta}}$,
with~$\bar{\Delta}=4\lambda^{3}+8\kappa\lambda^{2}+\lambda\left(  \kappa
^{2}+6\right)  +27\kappa$ from which we conclude that the exact scaling
solutions at the stationary points are stable, i.e. points $P_{3}^{\left(
\pm\right)  }\left(  D\right)  $ are attractors when $\left\{  \lambda
<0,~\lambda\left(  6+\lambda\left(  \kappa+\lambda\right)  \right)
>0,\sqrt{48+\frac{729}{\lambda^{2}}}\leq8\kappa+\frac{27}{\lambda+8\lambda
}\right\}  $ and~ $\left\{  \lambda>0,~\left(  27+\lambda\left(  8\kappa
+\sqrt{48+\frac{729}{\lambda^{2}}}\right)  \right)  <0~,-\frac{27}{8\lambda
}-\lambda-\frac{1}{8}\sqrt{48+\frac{729}{^{\lambda^{2}}}}\leq\kappa
<-\frac{6+\lambda^{2}}{\lambda}\right\}  $ . The latter regions are plotted in
Fig. \ref{fig9}. In Fig. \ref{fig10} the phase portrait of the field equations
in the space of variables $\left(  x,z\right)  $ is presented, while the
qualitative evolution of the effective equation of state parameter is
presented in Fig. \ref{fig11}. We observe that always $w_{eff}<-1$ and the
$w_{eff}$ cannot cross the phantom divide line. Thus this model is not
physically acceptable.

\subsubsection{Analysis at infinity}

For mathematical completeness we proceed with the determination of stationary
points at the infinity. In order to perform such an analysis we consider the
new coordinates%
\begin{equation}
x=\frac{\cos u\cos v}{\rho}~,~y=\frac{\sin u}{\rho}~,~z=\frac{\cos u\sin
v}{\rho},
\end{equation}
from where we find the equivalent dynamical system%
\begin{equation}
\rho^{\prime}=\frac{1}{2}\left(  1+\rho^{2}\right)  \left(  3\left(  \rho
^{2}-1\right)  +\lambda\rho\sqrt{3\left(  1-\rho^{2}\right)  }\cos v\right)
~,~
\end{equation}%
\begin{equation}
v^{\prime}=\frac{\sqrt{3}}{2\sqrt{1-\rho^{2}}}\left(  \kappa+\lambda+\rho
^{2}\left(  \lambda-\kappa\right)  \right)  \sin v~,~
\end{equation}
with equation of constraint $\rho^{2}+\cos\left(  2u\right)  =0$. The
stationary points with $\rho=0$, are the points with $\sin v=0$. Thus, in the
surface $\rho=0$, $\rho^{\prime}=\frac{3}{2}$, which means that the exact
solutions at the points are sources.

\section{Conclusions}

\label{sec5}

In this work we considered a two scalar field cosmological model for which the
two scalar fields are minimally coupled to gravity, but they have an
interaction term in the definition of their energy. Specifically, the dynamics
of the two scalar fields evolves in a two-dimensional space of constant
curvature. If the signature of the space is Euclidean and the scalar fields
have positive kinetic energy, then our gravitational Action Integral takes the
form of the Chiral model. However, in this analysis we considered the scalar
fields to have also negative energy density. This leads to the requirement
that the dynamics of the scalar fields evolve in a space of constant positive
or negative curvature with Lorentzian or Euclidean signature

We performed a detailed analysis on the dynamics of this specific cosmological
model in a spatially flat FLRW background space. In particular we determined
the stationary points and we investigated their stability, in order to study
the asymptotic behaviour of this cosmological theory and to understand the
cosmological evolution as also to investigate the existence of cosmological
solutions of special interest. For the completeness of our study we
investigated four different cases, which we called them models A, B, C and D.

Model A corresponds to the Chiral model, in which the two scalar fields have
positive kinetic energy. We found that the gravitational field equations admit
three different kinds of exact solutions which correspond to the two scaling
solutions of the quintessence (one scalar field model) while the third
solution is also a scaling solution wherein the two scalar fields contribute
to the cosmological evolution. The effective parameter for the equation of
state has lower bound at $-1,$ that is $w_{eff}\left(  A\right)  >-1$.

For model B, the second scalar field has a negative kinetic energy and can be
seen as the generalization of the quintom theory. Indeed, the quintom model is
recovered when the coupling parameter of the two scalar fields becomes zero.
The number of the stationary points are exactly the same as for Model A. Thus
in this case the parameter for the equation of state for the effective fluid
can cross the phantom divide lines, twice, which means that it can start from
a value larger than $-1$, then become smaller than $-1$ and at the end to
reach again a value larger than $-1$. That is exactly similar to the behaviour
of the equation of state parameter for the quintom model. It is important to
mention that in our numerical simulation we have not seen the appearance of
ghosts. Hence, that makes this specific case of special interest for further investigation.

Models C and D admit only two different sets of scaling solutions in the
finite region, while only model C admits an additional point at infinity. For
model C the equation of state parameter can cross the phantom divide line only
one time, but for model D the equation of state parameter is always lower than
$-1$, which means that the model D is not of physical interest.

From the above analysis we conclude that model B, which can be seen as a
generalization of the quintom model can describe some of the recent
observations and deserves further attention. In a forthcoming work we will
investigate the existence of additional exact and solutions for this specific model.

\begin{acknowledgments}
AP\ \& GL were funded by Agencia Nacional de Investigaci\'{o}n y Desarrollo -
ANID through the program FONDECYT Iniciaci\'{o}n grant no. 11180126.
Additionally, GL is supported by Vicerrector\'{\i}a de Investigaci\'{o}n y
Desarrollo Tecnol\'{o}gico at Universidad Catolica del Norte. AP thanks Prof.
P.G.L. Leach for his continuous support on the subject.
\end{acknowledgments}

\end{document}